%% file: master.tex
\documentclass[a4paper,11pt]{article}
\usepackage{pos}
\RequirePackage{wrapfig}
\RequirePackage{subcaption}
\RequirePackage{tikz}
\usepackage{color}

\title{Low-Temperature Hybrid Monte Carlo for Perylene and Corannulene: Stabilizing Determinant and Force}
\ShortTitle{Low-Temperature Hybrid Monte Carlo for Perylene and Corannulene}

\author*[a]{Petar Sinilkov}
\author[a]{Thomas Luu}
\author[b]{Johann Ostmeyer}
\author[a]{Finn L. Temmen}
\author[c]{Marcel Rodekamp}

\affiliation[a]{Institute for Advanced Simulation (IAS-4), \\
Forschungszentrum J\"{u}lich \\
Wilhelm-Johnen-Stra{\ss}e, 52428 J\"{u}lich, Germany}

\affiliation[b]{Helmholtz-Institut f\"{u}r Strahlen- und Kernphysik, Universit\"{a}t Bonn, \\
53115 Bonn, Germany}

\affiliation[c]{Institut f\"{u}r Theoretische Physik, Universit\"{a}t Regensburg, \\
93040 Regensburg, Germany}

\emailAdd{p.sinilkov@fz-juelich.de}
\emailAdd{t.luu@fz-juelich.de}
\emailAdd{ostmeyer@hiskp.uni-bonn.de}
\emailAdd{f.temmen@fz-juelich.de}
\emailAdd{marcel.rodekamp@ur.de}

\abstract{We demonstrate a real world physics study of the molecules perylene and corannulene at different doping levels, and calculating selected one-, two- (exciton) and three-body (trion) correlators. This has been done by performing Hubbard-model simulations near room temperature. We find, between charge neutral and doubly charged plateaus of perylene, an interaction-induced $\Delta Q = 1$ plateau whose width grows with interaction strength and in which the ground state is a spin doublet rather than a singlet. The algorithmic advancement enabling this study simultaneously stabilizes the determinant and force calculations to machine precision at room temperature, while scaling as well as standard determinant Quantum Monte Carlo simulations of these systems. Furthermore, our formalism retains the full time-displaced (2-point) Green's function, allowing us to construct any $n$-body correlator, including those that contain disconnected diagrams.}

\FullConference{The 43rd International Symposium on Lattice Field Theory (LATTICE2026)\\
 26 July - 1 August 2026\\
University of Maryland, College Park, USA\\}

\input{macros}

\begin{document}
\maketitle

\input{section/introduction}
\input{section/low-temperature-hmc}
\input{section/warm-molecules}
\input{section/conclusion}

\acknowledgments{
  P.S. was supported through MKW NRW under the funding code NW21-024-A. T.L., J.O., and F.L.T. are supported by the Deutsche Forschungsgemeinschaft (DFG, German Research Foundation) as part of the CRC 1639 NuMeriQS-project no.~511713970. We gratefully acknowledge the computing time granted by the JARA Vergabegremium and provided on the JARA Partition part of the supercomputer JURECA at Forschungszentrum Jülich.
}

\bibliographystyle{JHEP}
\bibliography{master}

\end{document}

%% file: macros.tex
\usepackage{xspace}
\usepackage{bbm}

\newcommand{\repoURL}{https://github.com/evanberkowitz/latex-base}

\newcommand{\secref}[1]{Sec.~\ref{sec:#1}}
\newcommand{\Secref}[1]{Section~\ref{sec:#1}}

\newcommand{\figref}[1]{Fig.~\ref{fig:#1}\xspace}
\newcommand{\Figref}[1]{Figure~\ref{fig:#1}\xspace}
\renewcommand{\eqref}[1]{(\ref{eq:#1})\xspace}

\newcommand{\Reference}[1]{Reference~\cite{#1}}

\tikzset{
  bond/.style        ={line width=1.3pt, black!62, line cap=round},
  frustrated/.style  ={line width=2.4pt, red!72!black, line cap=round},
  siteA/.style       ={circle, draw=black!78, fill=black!78,
                       inner sep=0pt, minimum size=6pt},
  siteB/.style       ={circle, draw=black!78, fill=white, line width=0.9pt,
                       inner sep=0pt, minimum size=6pt},
  ringfill/.style    ={fill=red!8},
}

\newcommand{\order}[1]{\ensuremath{\mathcal{O}\left(#1\right)}\xspace}

\newcommand{\abs}[1]{\ensuremath{\left| #1 \right|}\xspace}

\newcommand{\average}[1]{\ensuremath{\left\langle #1 \right\rangle}\xspace}

\let\builtinLaTeX\LaTeX
\def\LaTeX{\builtinLaTeX\xspace}

%% file: section/introduction.tex
\section{Introduction}\label{sec:intro}

Quantum Monte Carlo (QMC) studies of the Hubbard model carried out at large $\beta$ (inverse temperature) keep the finite temperature effects like the zero-temperature extrapolation and excited state decay in correlator analysis under control. Large $\beta$ also reaches near room temperature or below, at which experiments are performed, so results are easier to compare with measurements.

A recent Hybrid Monte Carlo (HMC) study of doped $\mathrm{C}_{20}\mathrm{H}_{12}$-perylene~\cite{perylene} computed the average charge $Q$ of the molecule as a function of the chemical potential $\mu$. In the non-interacting zero-temperature limit $Q(\mu)$ is a step function, with a sharp transition from the charge-zero to the charge-two state at the value of $\mu$ that first makes it favorable to add a pair of electrons. In contrast, their data, taken at $\beta = 4$, $6$ and $8$, showed a smooth transition. This is due to their ensembles being calculated at high temperature. Therefore, the transition structure is smeared out.

The obstacle to simply lowering the temperature is algorithmical rather than physical. Both the fermion determinant and the molecular-dynamics force of the HMC algorithm~\cite{hmc} are built by accumulating a long product of transfer matrices, and that product becomes catastrophically ill-conditioned as we increase $\beta$. At $U = 0$ its condition number is
\begin{equation}\label{eq:cond}
    \mathrm{cond}\Big(\prod_t M_t\Big) = e^{2\beta K},
\end{equation}
where $M_t$ is the transfer matrix of time slice $t$ and $K$ the spectral radius
of the hopping matrix. The dependence on $\beta$ is exponential, so double precision is
exhausted at a fairly modest inverse temperature.

Stabilizing the product of transfer matrices by repeated matrix decomposition is
standard practice in determinant quantum Monte Carlo (dQMC), and for the determinant
alone the problem has been alleviated in multiple studies~\cite{loh2005,Assaad2002,Bauer2020,kreit2026}.
The force also needs the equal-time Green's function at every time slice, the
same object dQMC stabilizes during its
sweeps~\cite{Assaad2002,Bauer2020}, but HMC needs it at every
molecular-dynamics step.  In \Reference{stablehmc} we showed how to obtain it
stably within HMC at \order{N_x^3 N_t} scale, which is what makes HMC at
$\beta \gtrsim 90$, i.e.\@ near room temperature, possible.

The intermediate quantities that make the force stable turn out to be the same
partial products from which the full time-displaced two-point function is
assembled.  The columns and rows of the propagator give the connected
contributions and its diagonal gives the disconnected ones, so once the force is
under control the entire propagator is available at no extra computational cost, and from it
any $n$-body two-point correlator can be built.

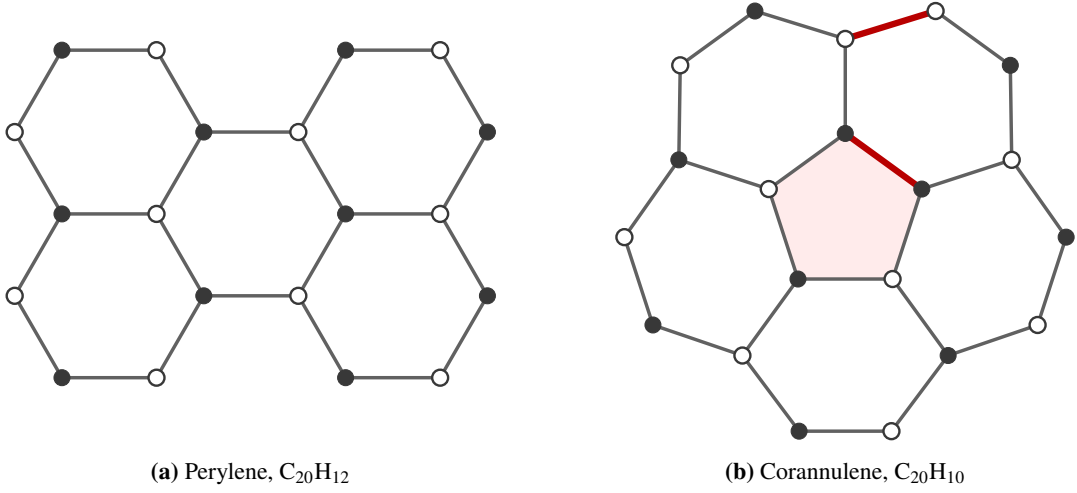
\begin{figure}[t]
    \centering
    \begin{subfigure}[b]{0.48\linewidth}
        \centering
        \begin{tikzpicture}[scale=1.25]
            \path (-2.55,-2.4) rectangle (2.55,2.4);
            \coordinate (A1)  at ( 1.000,  0.000);
            \coordinate (A2)  at (-0.500,  0.866);
            \coordinate (A3)  at (-0.500, -0.866);
            \coordinate (A4)  at ( 1.000,  1.732);
            \coordinate (A5)  at ( 1.000, -1.732);
            \coordinate (A6)  at ( 2.500,  0.866);
            \coordinate (A7)  at ( 2.500, -0.866);
            \coordinate (A8)  at (-2.000,  0.000);
            \coordinate (A9)  at (-2.000,  1.732);
            \coordinate (A10) at (-2.000, -1.732);
            \coordinate (B1)  at ( 0.500,  0.866);
            \coordinate (B2)  at ( 0.500, -0.866);
            \coordinate (B3)  at ( 2.000,  0.000);
            \coordinate (B4)  at (-1.000,  0.000);
            \coordinate (B5)  at (-1.000,  1.732);
            \coordinate (B6)  at (-1.000, -1.732);
            \coordinate (B7)  at ( 2.000,  1.732);
            \coordinate (B8)  at ( 2.000, -1.732);
            \coordinate (B9)  at (-2.500,  0.866);
            \coordinate (B10) at (-2.500, -0.866);
            \draw[bond] (A1)--(B1)--(A2)--(B4)--(A3)--(B2)--(A1);
            \draw[bond] (A6)--(B7)--(A4)--(B1);
            \draw[bond] (A1)--(B3)--(A6);
            \draw[bond] (A7)--(B3);
            \draw[bond] (B2)--(A5)--(B8)--(A7);
            \draw[bond] (A2)--(B5)--(A9)--(B9)--(A8)--(B4);
            \draw[bond] (A8)--(B10)--(A10)--(B6)--(A3);
            \foreach \n in {A1,A2,A3,A4,A5,A6,A7,A8,A9,A10}
                \node[siteA] at (\n) {};
            \foreach \n in {B1,B2,B3,B4,B5,B6,B7,B8,B9,B10}
                \node[siteB] at (\n) {};
        \end{tikzpicture}
        \caption{Perylene, $\mathrm{C}_{20}\mathrm{H}_{12}$}
        \label{fig:lattice-perylene}
    \end{subfigure}
    \hfill
    \begin{subfigure}[b]{0.48\linewidth}
        \centering
        \begin{tikzpicture}[scale=1.25]
            \path (-2.55,-2.4) rectangle (2.55,2.4);
            \foreach \k in {0,...,4} {
                \pgfmathsetmacro{\t}{90+72*\k}
                \coordinate (P\k)  at (\t:0.851);
                \coordinate (S\k)  at (\t:1.851);
                \coordinate (Ra\k) at ({\t+24}:2.349);
                \coordinate (Rb\k) at ({\t+48}:2.349);
            }
            \fill[ringfill] (P0)--(P1)--(P2)--(P3)--(P4)--cycle;
            \foreach \k [evaluate=\k as \kk using {int(mod(\k+1,5))}] in {0,...,4} {
                \draw[bond] (P\k)--(P\kk);
                \draw[bond] (P\k)--(S\k);
                \draw[bond] (S\k)--(Ra\k)--(Rb\k)--(S\kk);
            }
            \draw[frustrated] (P4)--(P0);
            \draw[frustrated] (Rb4)--(S0);
            \foreach \k in {0,2,4} {
                \node[siteA] at (P\k)  {};  \node[siteB] at (S\k)  {};
                \node[siteA] at (Ra\k) {};  \node[siteB] at (Rb\k) {};
            }
            \foreach \k in {1,3} {
                \node[siteB] at (P\k)  {};  \node[siteA] at (S\k)  {};
                \node[siteB] at (Ra\k) {};  \node[siteA] at (Rb\k) {};
            }
        \end{tikzpicture}
        \caption{Corannulene, $\mathrm{C}_{20}\mathrm{H}_{10}$}
        \label{fig:lattice-corannulene}
    \end{subfigure}
    \caption{%
      Tight-binding lattices of (a)~perylene and (b)~corannulene, one site per
      carbon $p_z$ orbital, $N_x = 20$ each.  Filled and open circles are the
      two sublattices of a two-coloring.  It succeeds for perylene, which is
      bipartite.  In corannulene the shaded pentagon and the fifteen-site
      perimeter are edge-disjoint odd cycles, so at least two bonds (red) must
      join sites of the same sublattice.  Corannulene is shown in projection.}
    \label{fig:molecules}
\end{figure}

For the first time we demonstrate the effectiveness of the algorithmic advances~\cite{stablehmc} to a real world physics study, namely the molecules perylene ($\mathrm{C}_{20}\mathrm{H}_{12}$)
and corannulene ($\mathrm{C}_{20}\mathrm{H}_{10}$) which are depicted in figure~\ref{fig:molecules}.
Polycyclic aromatic hydrocarbons such as these have long been proposed as carriers of the diffuse
interstellar bands~\cite{tielens2008}, but $\mathrm{C}_{60}^{+}$ remains the only
firm identification~\cite{campbell2015}, and corannulene, a bowl-shaped fragment
of $\mathrm{C}_{60}$, has been searched for without detection~\cite{pilleri2009};
such identifications need reliable spectra of the candidate molecules.  Perylene and corannulene are also of interest in organic light-emitting diodes and solar cells~\cite{sato1998,tang1986,chen2016,zoppi2011}.

\Secref{lt-hmc} fixes the Hubbard Hamiltonian and outlines the stabilized
algorithm and the correlators it gives access to.  \Secref{warm-molecules}
presents results for perylene and corannulene at $\beta = 90$, and
\secref{conclusion} summarizes and looks ahead.

%% file: section/low-temperature-hmc.tex
\section{Low-Temperature HMC}\label{sec:lt-hmc}

\subsection{The Hubbard model}\label{sec:theory}

We treat both molecules with the Hubbard model~\cite{hubbard1963}, where each carbon atom contributes one
site to the tight-binding lattice, giving $N_x = 20$ for both molecules, written in
the so-called spin basis as
\begin{equation}\label{eq:hubbard}
    H = -\kappa \!\!\sum_{\langle x,y\rangle,\, s=\uparrow,\downarrow}\!\!
          c_{x,s}^\dagger c_{y,s}
        - \frac{U}{2}\sum_{x}
          \left(n_{x,\uparrow} - n_{x,\downarrow}\right)^2
        - \mu\sum_{x}
          \left(n_{x,\uparrow} + n_{x,\downarrow}\right),
    \qquad n_{x,s} \equiv c_{x,s}^\dagger c_{x,s},
\end{equation}
where $c^\dagger$ and $c$ are the creation and annihilation operators, $\kappa$
is the hopping parameter, $U$ the on-site interaction and $\mu$ the chemical
potential.  The first sum runs over nearest-neighbor pairs $\langle x,y\rangle$
of carbon sites, one site per $p_z$ orbital, and $s$ over the two spins. The spin projection is defined as $S_z \equiv \tfrac{1}{2}\sum_x (n_{x,\uparrow} - n_{x,\downarrow})$ and its magnitude
$\abs{S_z}$. The last of these is what distinguishes a state carrying net spin
from a singlet:
in an ensemble that samples both signs of $S_z$ equally, $\average{S_z}$
vanishes while $\average{\abs{S_z}}$ does not.
Half filling corresponds to $\mu = 0$, and throughout we quote the molecule's total charge $Q$
relative to it, so that $\average{Q} = 0$ at $\mu = 0$ and the plateaus of
\secref{perylene} count electrons added to the neutral molecule.

Throughout we work in units of the hopping, so that $\beta$, $U$ and $\mu$ are
all dimensionless.  Taking the value $\kappa = 2.4$~eV for the
$\pi$-electron hopping of a carbon lattice~\cite{giri2018}, the inverse temperature
$\beta = 90\,(\approx\,309\,\mathrm{K})$ at which we run is near room temperature, and an order of
magnitude colder than the $\beta = 4$--$8\,(\approx\,6960$--$3480\,\mathrm{K})$ ensembles of \Reference{perylene}.

\subsection{Stable HMC and \texorpdfstring{$n$}{n}-body correlators}
\label{sec:cor-func}

We sample \eqref{hubbard} with HMC~\cite{hmc}, discretizing Euclidean time into $N_t$ slices of extent $\delta = \beta/N_t$ and decoupling the on-site interaction with a Hubbard--Stratonovich field so that the fermions can be integrated out exactly~\cite{bss1981}. Every ensemble below was generated with the stabilized algorithm of \Reference{stablehmc}, to which we refer for the construction, the choice of decomposition and the benchmarks. The entire propagator is available on every configuration at no extra cost, and with it any $n$-body two-point correlator.

We measure one-, two- and three-body correlators on the same ensembles.  Writing
$c^{\dagger}_{k\pm,s}$ for the creation operator of the single-particle
eigenmode $k\pm$ of the hopping matrix, with spin
$s \in \{\uparrow,\downarrow\}$ and $\pm$ labeling the two branches of the
spectrum about half filling, the interpolating operators are
\begin{align}
    \mathcal{O}^{(1)}_{k+,\uparrow} &= c^{\dagger}_{k+,\uparrow},
        \label{eq:op1}\\
    \mathcal{O}^{(2)}_{k+,l-} &=
        \frac{1}{\sqrt{2}}\left(c^{\dagger}_{k+,\uparrow} c_{l-,\uparrow}
            - c^{\dagger}_{k+,\downarrow} c_{l-,\downarrow}\right),
        \label{eq:op2}\\
    \mathcal{O}^{(3)}_{k+,k+,k-} &=
        c^{\dagger}_{k+,\uparrow} c^{\dagger}_{k+,\downarrow}
        c_{k-,\downarrow},
        \label{eq:op3}
\end{align}
with $C^{(n)}(\tau) = \average{\mathcal{O}^{(n)}(\tau)\,
\mathcal{O}^{(n)\dagger}(0)}$.  Operator~\eqref{op1} adds a single electron to
band $+$, \eqref{op2} promotes an electron from band $-$ to band $+$ as the
$S_z = 0$ member of a spin triplet, and \eqref{op3} adds two electrons to band
$+$ while removing one from band $-$; we call the two-body channel the exciton
and the three-body channel the trion.  The other members of the triplet, such as
$c^{\dagger}_{k+,\uparrow} c_{l-,\downarrow}$, are degenerate with
\eqref{op2}.
These one-body modes are adequate interpolating operators: hopping and on-site
interaction share the symmetry of the molecule, so each operator carries quantum
numbers the interacting theory conserves, and its overlap with the target state
is exact at $U = 0$.

%% file: section/warm-molecules.tex
\section{Warm Molecules}\label{sec:warm-molecules}

\subsection{Setup and simulation parameters}\label{sec:setup}
Perylene is built entirely from six-membered rings, so its lattice graph
contains only even cycles and is bipartite.  Corannulene has a five-membered
ring at its center, an odd cycle, and its lattice is therefore not bipartite.
Bipartiteness is what protects the half-filled Hubbard model from the sign
problem.  Corannulene loses that protection, although at the couplings used here
its average sign stays at unity near half filling (\figref{corannulene}).

Where the fermion determinant can be negative we sample its absolute value and
reweight,
\begin{equation}\label{eq:reweight}
    \average{\mathcal{O}}
      = \frac{\average{\mathcal{O}\,\Sigma}_{\abs{\cdot}}}
             {\average{\Sigma}_{\abs{\cdot}}},
\end{equation}
with $\average{\cdot}_{\abs{\cdot}}$ the average in that ensemble and
$\Sigma = \pm 1$ the sign of the fermion determinant.  The denominator is the
average sign $\average{\Sigma}$, which we quote alongside every
charge scan, since it is a measure of how much statistical power a
given ensemble retains: the statistical error of a reweighted observable grows
as $1/(\average{\Sigma}\sqrt{N_{\mathrm{cfg}}})$, so the number of configurations
needed grows as $\average{\Sigma}^{-2}$.
All ensembles below have $N_{\mathrm{cfg}} \approx 10^4$ configurations at each value
of $\mu$.

\subsection{Perylene}\label{sec:perylene}

\begin{figure}[t]
    \centering
    \includegraphics[width=0.85\linewidth]{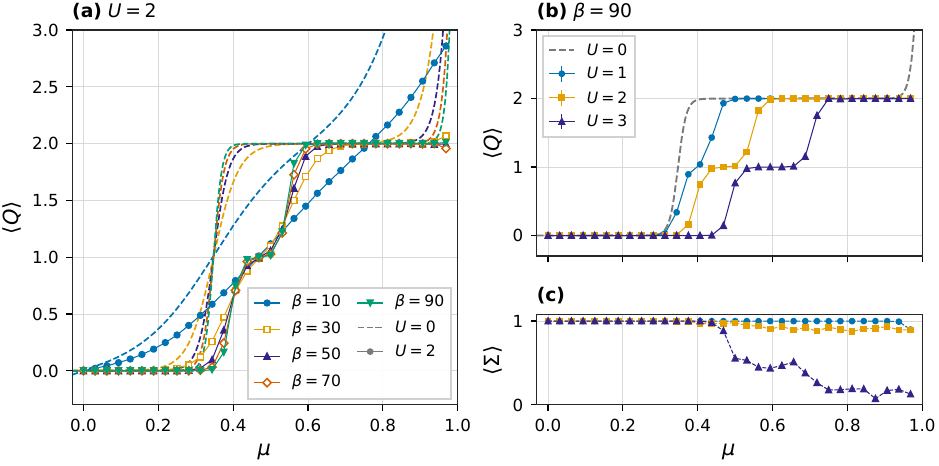}
    \caption{%
      Average charge of perylene against chemical potential, at fixed step
      size $\delta = \beta/N_t \approx 0.18$.  (a)~Temperature dependence at
      $U = 2$: interacting data (markers) for $\beta = 10$, 30, 50, 70 and 90,
      with the non-interacting result at each $\beta$ (dashed).  The smooth
      ramp of the warm ensembles sharpens into a step function as $\beta$ grows.
      (b)~Interaction dependence at $\beta = 90$: the $\Delta Q = 1$ plateau
      appears only for $U > 0$ and widens with $U$.  (c)~The corresponding
      average sign, which stays close to unity for $U = 1$ and 2 but collapses
      for $U = 3$.
      }
    \label{fig:perylene-charge}
\end{figure}

Simulating at $\beta = 90$ recovers the transition structure. Figure~\ref{fig:perylene-charge}(a) repeats the charge scan at
$\beta = 10$, $30$, $50$, $70$ and $90$: as the ensembles cool, the smooth
ramp of the warm data steepens into sharp transitions.

Between the $Q = 0$ and $Q = 2$ plateaus a third plateau appears at
$Q = 1$.  It is absent from the non-interacting ($U = 0$) result at any temperature; it exists only once the interaction is switched on.
At $U=0$ both spin states of the lowest unoccupied molecular orbital (LUMO) have degenerate energy.
In consequence, the addition of an electron of either spin to the LUMO level comes at equal energy cost and happens simultaneously.
The on-site repulsion, in contrast, makes the second electron more energetically expensive, splitting the transition in two and leaving a region in which the molecule holds a single unpaired electron. In this single electron region the spin symmetry is spontaneously broken.

Figure~\ref{fig:perylene-charge}(b) scans $U$ at
fixed $\beta = 90$, and the plateau widens monotonically with the
interaction strength while the whole structure shifts to larger $\mu$. Its
width is the charging energy of LUMO i.e. the energy cost of adding an electron. The average sign shown in Fig.~\ref{fig:perylene-charge}(c) stays close to unity for $U = 1$ and $U = 2$, but at $U = 3$ it collapses across the transition region, so the ensembles that show the effect most clearly are the ones with the least statistical power.

\begin{figure}[t]
    \centering
    \includegraphics[width=0.8\linewidth]{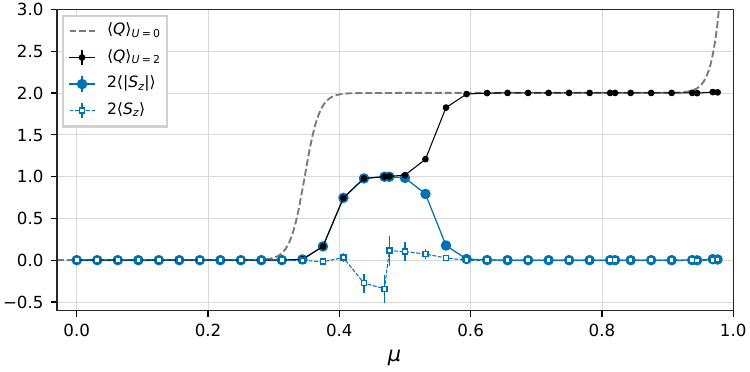}
    \caption{%
      Spin structure of the $Q = 1$ region of perylene at $U = 2$,
      $\beta = 90$, $N_t = 512$.  Across the plateau $2\average{\abs{S_z}}$
      rises to unity while $2\average{S_z}$ stays consistent with zero,
      identifying the ground state as a spin doublet rather than a singlet.  The
      charge at $U = 2$ and $U = 0$ is shown for reference.}
    \label{fig:perylene-spin}
\end{figure}

An odd total charge cannot be accommodated by a spin singlet, so throughout the
$Q = 1$ plateau the ground state is a spin doublet.
\Figref{perylene-spin} makes this visible at $U = 2$: across the plateau
$\average{\abs{S_z}}$ rises to $\tfrac{1}{2}$ while $\average{S_z}$ remains consistent
with zero, the signature of an ensemble sampling both polarizations of a
genuinely spin-$\tfrac{1}{2}$ state, and outside the plateau both observables
vanish.

\begin{figure}[t]
    \centering
    \includegraphics[width=0.9\linewidth]{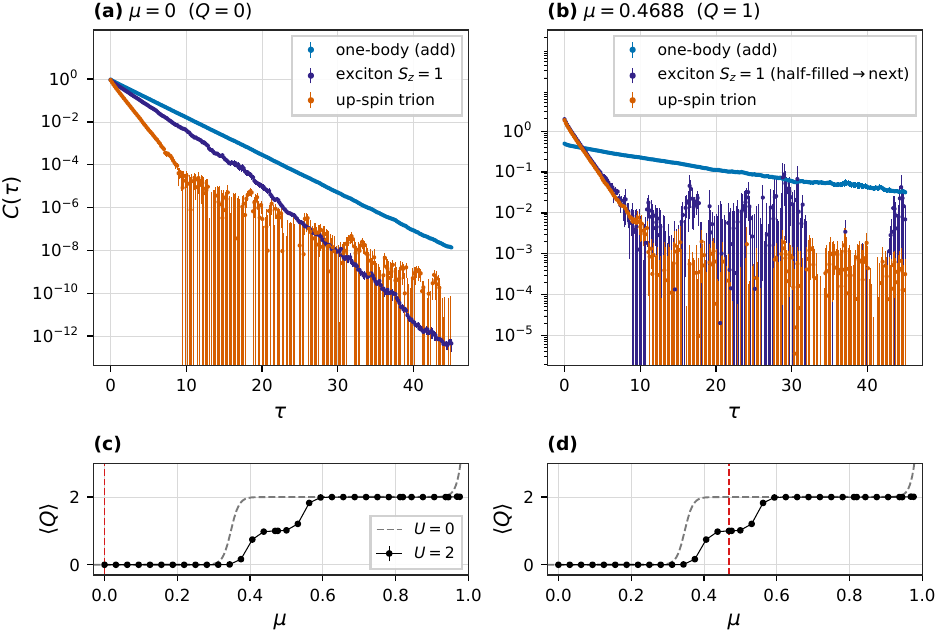}
    \caption{%
      One-, two- and three-body correlation functions of perylene at $U = 2$,
      $\beta = 90$, $N_t = 512$: (a)~at half filling and (b)~in the middle of
      the $Q = 1$ plateau, at $\mu = 0.4688$.  Panels (c) and (d) mark the
      chemical potential of each set on the charge curve.  At $\mu = 0$ the
      exciton promotes an electron across the gap; in the plateau it promotes
      the unpaired electron to the next level.  The exciton curve, labelled
      $S_z = 1$, is the degenerate triplet partner
      $c^{\dagger}_{k+,\uparrow} c_{l-,\downarrow}$ of \eqref{op2}.}
    \label{fig:perylene-correlators}
\end{figure}

The full propagator lets us look inside both regions rather than only at their
charge.  \Figref{perylene-correlators} compares one-, two- and three-body
correlators at half filling with the same correlators in the middle of the
$Q = 1$ plateau.  At $\mu = 0$ the one- and two-body correlators decay cleanly,
while the three-body one becomes noisy at larger $\tau$.  Inside the plateau the
one-body correlator barely decays, because adding or removing an electron there
costs only the distance to the nearer charge transition, and the two- and
three-body channels are not distinguishable.

\subsection{Corannulene}\label{sec:corannulene}

\begin{figure}[t]
    \centering
    \includegraphics[width=0.85\linewidth]{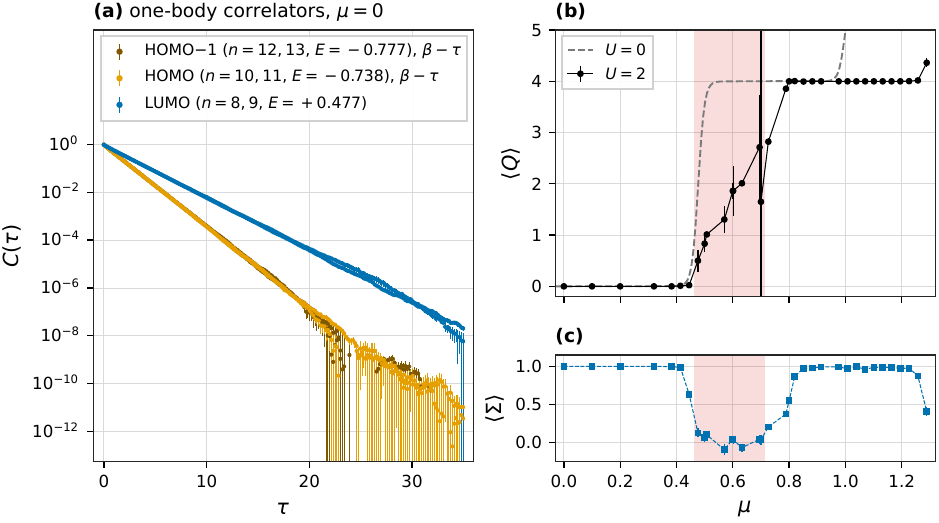}
    \caption{%
      Corannulene at $U = 2$, $\beta = 90$, $N_t = 512$.  (a)~One-body
      correlation functions at $\mu = 0$, one color per degenerate pair of
      levels.  For the empty LUMO the added electron propagates cleanly forward
      in $\tau$; for the filled HOMO and HOMO$-1$ the forward branch is
      Pauli-blocked, and the signal is the backward branch, where an electron is removed, plotted against $\beta - \tau$. The slopes differ because
      the spectrum of the non-bipartite lattice is not symmetric about half
      filling. (b)~Average charge and (c)~average sign against chemical
      potential: the sign is unity at half filling and collapses exactly where
      the charge transition happens.  Shading marks a severe sign problem,
      $\average{\Sigma} \le 0.2$.}
    \label{fig:corannulene}
\end{figure}

Figures~\ref{fig:corannulene}(b) and~(c) show the charge and average sign of
corannulene at $U = 2$, $\beta = 90$.  The non-interacting reference steps from $Q = 0$ to $Q = 4$ and
the interacting data climb through intermediate values~\cite{baumgarten1995,ayalon1994}, hinting at the
substructure resolved cleanly in perylene.  We cannot confirm it across
$0.45 \lesssim \mu \lesssim 0.8$ because the average sign in this region is consistent with zero, the error bars
are uncontrolled, and no plateau can be identified.

At half filling the average sign is unity, and Fig.~\ref{fig:corannulene}(a)
shows instead the single-particle structure around the gap.
Adding an electron to the empty LUMO gives a clean forward decay in $\tau$; for
the highest occupied molecular orbital (HOMO) and HOMO$-1$ the forward branch is Pauli-blocked, and the clean
signal is the backward branch towards $\tau = \beta$, where an electron is
removed.  The two decay rates differ because the spectrum of the non-bipartite
lattice is not symmetric about half filling: in the non-interacting spectrum the
LUMO lies at $+0.477$ and the HOMO at $-0.738$.

%% file: section/conclusion.tex
\section{Summary and Outlook}\label{sec:conclusion}

Stabilizing the determinant and the force together makes HMC usable at inverse
temperatures where it previously broke down, without changing its
\order{N_x^3 N_t} scaling.  The zero-temperature limit becomes accessible to a
controlled extrapolation rather than an optimistic one, and the full
time-displaced propagator falls out of the same partial products that stabilize
the force, so every $n$-body two-point correlator, connected and disconnected
alike, can be measured on the ensembles we already have.
We are currently analysing these correlators in detail using the Truncated Hankel Correlator (THC) method~\cite{Ostmeyer:2025igc} in order to obtain the low-energy excitation spectrum.

Applying this to perylene and corannulene at $\beta = 90$, we recover the charge
transition that higher temperature ensembles cannot resolve, and we find between its endpoints a
$\Delta Q = 1$ plateau that exists only in the interacting theory and widens
with $U$.  In that window the ground state is not a singlet: $\average{\abs{S_z}}$
rises to $\tfrac{1}{2}$ while $\average{S_z}$ stays at zero.  We have shown one-, two-
and three-body correlators both at half filling and inside the new region. For corannulene the sign problem limits the analysis across the charge transition, while at half filling the correlators show the asymmetric spectrum of the non-bipartite lattice.

The Hubbard interaction is purely on-site, although for twenty $\pi$ electrons this close together the Coulomb tail between neighboring sites is not obviously negligible. The
Pariser--Parr--Pople model~\cite{fabian2025} with Mataga--Nishimoto interpolation~\cite{mataga1957} keeps that tail, replacing the single parameter $U$ with a distance-dependent interaction. Work on perylene and corannulene in this model is under way.

\section*{Declaration of AI-assisted tools}

During the preparation of this manuscript, the authors used Claude Opus 5.5 to improve the clarity of selected paragraphs and the data visualization. The authors reviewed, edited, and approved all text and take full responsibility for the content.